\documentclass[apj]{emulateapj}
\newcommand{\beq}{\begin{equation}}
\newcommand{\eeq}{\end{equation}}

\newcommand{\citei}[1]{\citeauthor{#1} \citeyear{#1}}

\newcommand{\kms}{km ${\rm s^{-1}}$~}
\slugcomment{Accepted by The Astrophysical Journal}
\begin{document}

\title{Ongoing Galactic Accretion: Simulations and Observations of Condensed Gas in Hot Halos}

\author{J.~E.~G.~Peek\altaffilmark{1}, M.~E.~Putman\altaffilmark{2}, Jesper Sommer-Larsen\altaffilmark{3,4}}

\altaffiltext{1}{Department of Astronomy, University of California, Berkeley, CA 94720}
\altaffiltext{2}{Department of Astronomy, University of Michigan, 500 Church Street, Ann Arbor, MI 48109}
\altaffiltext{3}{Dark Cosmology Centre, Niels Bohr Institute, Juliane Maries Vej 30, DK-2100 Copenhagen, Denmark}
\altaffiltext{4}{Institute of Astronomy, University of Tokyo, Osawa 2-21-1,
Mitaka, Tokyo, 181-0015, Japan}

\begin{abstract}
Ongoing accretion onto galactic disks has been recently theorized to progress via the unstable cooling of the baryonic halo into condensed clouds. These clouds have been identified as analogous to the High-Velocity Clouds (HVCs) observed in HI in our Galaxy. Here we compare the distribution of HVCs observed around our own Galaxy and extra-planar gas around the Andromeda galaxy to these possible HVC analogs in a simulation of galaxy formation that naturally generates these condensed clouds. We find a very good correspondence between these observations and the simulation, in terms of number, angular size, velocity distribution, overall flux and flux distribution of the clouds. We show that condensed cloud accretion only accounts for $\sim 0.2 M_{\odot}$ / year of the current overall Galactic accretion in the simulations. We also find that the simulated halo clouds accelerate and become more massive as they fall toward the disk. The parameter space of the simulated clouds is consistent with all of the observed HVC complexes that have distance constraints, except the Magellanic Stream which is known to have a different origin. We also find that nearly half of these simulated halo clouds would be indistinguishable from lower-velocity gas and that this effect is strongest further from the disk of the galaxy, thus indicating a possible missing population of HVCs. These results indicate that the majority of HVCs are consistent with being infalling, condensed clouds that are a remnant of Galaxy formation.

\end{abstract}


\section{Introduction}

Galaxies have long been thought to form from the cooling of shock-heated primordial material in galactic halos (e.g.  \citei{WR78}, \citei{WF91}). These galaxy formation models, wherein all of the gas inside some time-dependent ``cooling radius'' monolithically collapses, cannot generate the continuous accretion needed to reproduce the Galaxy's star-formation history (e.g. \citei{RP2000}). \citet{MB2004} (hereafter MB04) showed that the process of the formation of galaxies could be driven by a multi-phase cooling process, in which not all of the gas within the cooling radius simply collapses (see also \citei{kaufmann07} and \citei{SL06}), but instead the hot ($\sim 10^6$ K) shock-heated gas condenses out into warm ($\sim 10^4$ K) clouds which fall inwards and collide, contributing to the formation of the galaxy. As time goes on, clouds condense at lower densities, corresponding to greater distances from the galactic center, and `rain' inwards towards the disk. This formulation, wherein accretion is more continuous, avoids the ``overcooling problem'' that plagued monolithic collapse models. MB04 identified these condensations as being analogous to observed High-Velocity Clouds (HVCs). This raised the exciting possibility that observations of HVCs can be used as a direct measure of the rate of our Galaxy's growth. We now further consider that we may be able to study galaxy formation quantitatively by studying various HVC observables and mapping them to the various parameters of galaxy formation simulations. Such a course of study might help disentangle the large scale processes (cooling, merging, outflows, etc.) that shape today's galaxy population.

HVCs have been known for over 40 years \citep{Muller1963}, having been first observed in the 21-cm hyperfine transition of neutral hydrogen. They are observed to be moving at hundreds of \kms, beyond what would be expected on the basis of Galactic rotation, with an overall sense of infall. Large maps have been made of the Galactic sky in HI (e.g. \citei{putman2002}, \citei{kalberla2005}), providing a good understanding of the gross observational characteristics of the HI HVCs. HVCs have also been observed in other atomic transitions (e.g \citei{sembach2003}, \citei{tripp03}), ionized gas (e.g. \citei{putman2003}, \citei{Tufte2002}) and there are tentative detections of dust as well \citep{M-D05}. These observations are much more sparse than the HI observations, and do not give us a sense of the distribution of the entire HVCs population, though they provide some useful fiducial characteristics, such as metallicity \citep{vww04} and rough estimates of ionization fraction \citep{Tufte2004}. Though the halo has been shown to have significant high ions toward some HVC sightlines \citep{fox06}, the majority of H$\alpha$ observations show limited ionization fractions in HVCs around 20\%. 

The origin and physical character of HVCs is still not known. The Magellanic Stream, a complex of HVCs trailing the Magellanic Clouds, is certainly due to the interaction of the Galaxy and the Magellanic Clouds and may have come from some combination of ram-pressure stripping by the gaseous halo and tidal effects \citep{putman2003}. Some Intermediate-Velocity Clouds (IVCs) have been shown to be very nearby ($\sim 1$ kpc from the Galactic disk) and have near-solar metallicities, and are therefore thought to be part of a Galactic fountain, a formation scenario in which gas is kicked out of the disk and rains back down. The rest of the population of HVCs are thought either to be satellite debris akin to the Magellanic Stream or part of the cooling formation process. In the past both Galactic fountain models (e.g. \citei{deAvillez2000}) and Local Group models, wherein HVCs are dark-matter dominated clouds hundreds of kpcs from the Galaxy (e.g. \citei{Blitz1999}), have also been invoked to explain HVCs. Neither of these models, though, is currently thought to be a successful explanation of the bulk of HVCs. Many HVCs have been shown to have distances beyond a few kpc (e.g. \citei{Thom2006}, \citei{Wakker2001}), and low metallicities (for a review, \citei{vww04}), which conflicts with the fountain model predictions of nearby, disk-like gas. Observations of Andromeda (\citei{Thilker2004}, \citei{Westmeier2005}) have excluded massive HVCs at great distances from that galaxy, which are a key prediction of Local Group models.

In an effort to understand the significance and scope of a Galactic gaseous halo, \citet{SL06} (hereafter S-L06) ran numerical simulations of galaxy formation in the $\Lambda$CDM cosmology. These simulations showed that in Milky-Way-sized galaxies, significant baryonic mass resides in the hot halo. S-L06 also showed that infalling condensations are a natural consequence of such halos, and that it is possible to compare these clouds to observed HVCs. \citet{Putman06} found the total mass in HVCs is consistent with the mass in clouds found by S-L06.

In this paper we study the condensations found in S-L06 as analogous to HVCs and compare them to the HI HVC population in our Galaxy in terms of their number, flux, spatial and velocity distribution. Since we do not expect the resolution of these simulations to be high enough to resolve the masses of all individual HVCs, we instead compare coherent groups of to HVCs to groups of simulated halo clouds. We will also compare the simulated condensations to the extra-planar gas observed around the Andromeda galaxy, which provides the opportunity to compare the projected distance of the clouds, albeit with much lower resolution and sensitivity. We show that these condensations are indeed consistent with the overall HVC population and examine the simulation population for physical characteristics that may inform our future study of HVCs and ongoing galaxy formation.

The structure of the paper is as follows: In \S \ref{obs} we discuss observations of HVCs around the Milky Way and and extragalactic HVCs around the Andromeda galaxy, in \S \ref{sims} we discuss the details of the S-L06 simulations and the halo clouds found in the simulations, in \S \ref{cplxing} we describe the process by which we self-consistently associate clouds in complexes, in \S \ref{anal} we compare the observational data with the simulation, in \S \ref{disc} we discuss the implications of a cooling-formation origin for HVC observables and we conclude in \S \ref{conc}. 

\section{Observations}\label{obs}

\subsection{Selecting HVCs from Galactic Observations}\label{selhvc}

As we are interested in the global properties of HVCs rather than their minutiae, which are beyond the resolution of the simulation, we wish to use a full sky survey with consistent resolution and nomenclature. For this reason we use the updated Wakker and Van Woerden catalog (\citei{WVW1991}, hereafter the WvWcat). This catalog has the velocities, fluxes and positions of more than 600 clouds; \citet{Wakker2004} uses this same sample, and a more detailed description can be found therein. 

The WvWcat includes all clouds that have historically been catalogued as HVCs, and therefore there is some inconsistency in their kinematic selection.  
For consistency we exclude those clouds that do not fit a deviation velocity criteria and a classical Local Standard of Rest (LSR) velocity criteria.
The deviation velocity,  $v_{\rm dev}$, is a measure of how the velocity of a cloud deviates from a relatively simple model of the Galactic disk. The model includes solid-body rotation in the Galactic center, a flat rotation curve at larger R and a flared disk (see \citei{Wakker2004} for details).  This is a useful criteria, as acceptable Galactic Standard of Rest (GSR) velocities
vary strongly as a function of angle on the sky, either requiring very harsh cuts ($|v_{\rm GSR}| > 220$ \kms) or including a large amount of gas plainly part of the disk in the quadrants I and IV.
Local Standard of Rest (LSR) velocities (which is to say heliocentric radial velocities corrected for the Sun's peculiar motion as compared to the mean motion of stars in the stellar vicinity) have been used historically to define the difference between HVCs and other Galactic clouds.   We use the criteria $|v_{\rm dev}| > 60$ \kms and $|v_{\rm LSR}| > 90$ \kms, both to excerpt the disk and to fulfill the classical definition of HVCs. These criteria are applied to observed HVCs and simulated halo clouds alike. The $|v_{\rm LSR}| > 90$ \kms has a very limited effect upon the selected complexes once the  $|v_{\rm dev}| > 60$ \kms is applied and our results do not significantly depend upon whether the $|v_{\rm LSR}| > 90$ \kms is applied.

In addition to the velocity selection described above, we exclude clouds which have origins that are not captured by the simulation, namely the Magellanic Stream and the Leading Arm of the Magellanic System. The Magellanic Stream and Leading Arm originated in the Magellanic Clouds (e.g. \citei{Bruns05}) and though the simulation could capture a similar dramatic disruption event of an accreting satellite, we find no such event in the simulated halo clouds (see section \ref{selsim}). After excluding these clouds and applying the velocity criteria, we are left with 412 clouds.

The majority of our remaining 412 clouds in the WvWcat are grouped into ``complexes'', or a large group of clouds in the same region of the sky with similar velocities. 
These complexes contain the majority of the HI flux of HVCs and, given the resolution of the simulations, form a useful comparison set.
Unfortunately, these complexes were defined in a historical and qualitative manner that is impossible to reproduce for the simulated dataset.
We therefore apply a quantitative, consistent method for generating complexes in both the simulated and observed data sets as described in section \ref{cplxing}. 

\subsection{Extragalactic HVCs around the Andromeda Galaxy}

In addition to Galactic observations of HVCs, extra-planar gas, seemingly analogous to HVC complexes, has been discovered in other galaxies. In particular, recent observations of Andromeda (M31) have shown a large number of distinct extra-planar clouds \citep{Thilker2004} (hereafter T04). To avoid confusion, we will refer to these clouds that reside in the halos of other galaxies as extragalactic HVCs. The T04 observations are the most comprehensive study of extragalactic HVCs and have enough sensitivity to detect a large fraction of condensed clouds analogous to those from the S-L06 simulations. These observations have an advantage over Galactic observations in that they have a projected galacto-centric radial distance for each cloud, as well as a relatively accurate distance and therefore HI mass for the clouds. At a distance of 775 kpc Andromeda is the nearest spiral galaxy to our own Milky Way. With a mass comparable to that of the Milky Way \citep{SBB06}, and without evidence of recent major mergers, we expect Andromeda to have a relatively similar recent formation history, and therefore a similar population of halo clouds, to our own Galaxy. The T04 observations cover a 94 kpc x 94 kpc square at Andromeda which, though smaller than the simulation domain, overlaps with the bulk of the simulated clouds. The resolution is 2 kpc with capacity to detect clouds down to a few $\times 10^{5} M_\odot$ of HI, depending upon their size, which is comparable to the mass resolution of the simulation. These observations (along with \citei{Westmeier2005}) show that there exists a significant population of extragalactic HVCs within 50 kpc of Andromeda's disk, with masses ranging from the sensitivity limit up to $10^7 M_{\odot}$, and that there are not extragalactic HVCs with masses $ \ge 10^6 M_\odot$ outside $R = 50$ kpc; these are significant constraints to which we can compare the simulations.

\section{Simulations}\label{sims}
The code used for the simulations is a significantly improved version of
the TreeSPH code, which has been used previously for galaxy formation 
simulations \citep{SLGP03}.
The main improvements over the previous version are:
(1) The ``conservative'' entropy 
equation solving scheme suggested by \citet{SH02} has been adopted. 
(2) Non-instantaneous gas recycling and chemical evolution, tracing
10 elements (H, He, C, N, O, Mg, Si, S, Ca and Fe), has been incorporated
in the code following \citet{Lia02a} and \citet{Lia02b}; the algorithm includes 
supernov\ae\ of type II and type Ia, and mass loss from stars of all masses.
(3) Atomic radiative cooling depending both on the metal abundance
of the gas and on the meta--galactic UV field, modeled after \citet{HM96} is invoked, as well as simplified treatment
of radiative transfer, switching off the UV field where the gas
becomes optically thick to Lyman limit photons on scales of $\sim$ 1~kpc.

\citet{SL06} selected a Milky-Way-like galaxy from a
cosmological simulation and simulated its gaseous halo 
at extremely high resolution. The purpose of the experiment
was to establish how large a mass fraction of the hot gas halos,
shown to contribute significantly to the baryonic mass budget of such
galaxies, condense into ``warm'' ($T$$\sim$$10^4$ K) clouds by
thermal instability. The result of the experiment was that only a
few percent of the hot gas mass forms warm clouds; it was
suggested that these clouds would be the equivalent of the HVCs. For
the purpose of this paper, specifically addressing the properties
of these warm clouds, an enlarged version of the above simulation 
was performed.

The base of the experiment was a 3.2$\times$10$^5$ particle, fully
cosmological simulation of a disk galaxy, which at $z$=0 has
a characteristic circular velocity of $V_c$=224 km/s, very similar to
that of the Milky Way. At $t$=10.0 Gyr ($z$$\sim$0.3), all gas particles 
within 250 kpc galacto-centric distance are split in eight particles
of mass 1/8th the original value and gravity softening length (inverse gravity force
resolution) 1/2 of the original value.
The simulation is then continued for 200 Myr, and then all gas particles 
within 100 kpc galacto-centric distance are again split in eight particles
of mass 1/8 and gravity softening length 1/2 of the previous values.
At this point the simulation totals 1.1$\times$10$^6$ particles, of
which 8.6$\times$10$^5$ are gas particles. The warm/hot gas in the inner
100 kpc of the galaxy halo is then resolved with particles of mass
$m_{\rm{gas}}$=11700 M$_{\odot}$ and gravity softening length
128 pc. As discussed by \citet{SL06} this enables the simulation
to resolve HVCs down to masses of $\sim 3 \times 10^{5} M_\odot$ within
100 kpc galacto-centric distance. The (by now) very high resolution simulation
is run for 500 Myr, until $t$=10.7 Gyr, at which point it was terminated due
to the heavy computational load. By  $t$=10.3 Gyr more than 100 condensed halo clouds
have formed, with the number increasing slightly over the next 300 Myr.
All HVCs are found at galacto-centric distances less than 100 kpc, no
HVCs were found in the region 100$<$$r$$\la$250 kpc down to
the mass resolution of $\sim 3 \times 10^{6} M_\odot$ in this region of the halo (Sommer-Larsen 2006). We note that the necessary condition for onset of thermal instability, viz. $\tau_{\lambda}<\tau_{\rm{cool}}$, is satisfied everywhere in the hot halo gas ($\tau_{\lambda}$ is the sound crossing time, which is taken to be $\sim 2h_{\rm{SPH}}/c_s$, where $h_{\rm{SPH}}$ is the local SPH smoothing length and $c_s$ is the sound speed; $\tau_{\rm{cool}}=\frac{E}{\dot{E}}$ is the timescale for radiative cooling). 

A number of physical processes can prevent the formation of warm
($T\sim10^4$ K) clouds out of ($T\sim10^6$ K) gas by thermal instability,
or lead to the destruction of warm clouds moving in such hot gas.
In particular, as discussed by MB04, these include thermal conduction,
Kelvin-Helmhotz (KH) instability and conductive evaporation. This will
impose a lower limit on the warm cloud mass, below which the above
processes significantly affect the clouds before they are accreted onto
the disk of the galaxy. As the number of SPH particles in the warm
clouds are typically $\la 500$ (section \ref{selsim}) and the simulation was
run with zero thermal conductivity, it is important to demonstrate
that the outcome of the simulation is not affected by numerical
resolution \citep{SL06} or lack of realistic modelling of
the physics involved. Moreover, \citet{Agertz06} showed that the
SPH method may have fundemental problems in modelling the KH
instability. On the other hand,
\citet{Vietri97} showed that KH cloud-breakup by the dominant
``champagne effect'' is highly suppressed by radiative cooling, which is included
in the present simulations, but not in the numerical experiments performed
by \citet{Agertz06}.

As discussed in \S \ref{physchar} and \S \ref{accretion}, the total mass of the
warm cloud system in the simulation is $\sim 10^8$ M$_{\odot}$,
and the accretion rate onto the disk is $\sim 0.2$ M$_{\odot}$/yr.
This gives a characteristic accretion timescale of $\sim 5\times10^8$ yr.
For such an accretion timescale, it follows from the work of MB04 that
warm clouds more massive than $\sim 10^4$ M$_{\odot}$ will not be
affected by thermal conduction/evaporation or KH breakup. As discussed
in section  \S \ref{physchar} the masses of the warm clouds identified in the
present simulation lie in the range $\sim 10^5 - 5 \times
10^6$ M$_{\odot}$, so even at much higher numerical resolution and with
thermal conduction included, warm clouds in the above mass range
should neither be prevented from forming nor suffer significant
destruction.

Based on the evolution of the underlying, lower mass resolution
simulation from t $\sim$ 0 to t=10 Gyr it appears that some of the halo clouds may
have been seeded by the remains of ``cold accretion'' events taking
place much earlier in the history of the galaxy, as discussed by
SL06. However, given the much lower resolution of the main
underlying galaxy simulation, starting at $z_i$=39 and running to
$t$=10.0 Gyr, it is not possible to give a detailed discussion of
the origin of the halo clouds on the basis of the present simulations. These constraints
also imply that the results obtained in this paper should,
in general, be regarded as preliminary. Eventually, simulations starting
at early times and of yet higher resolution should be undertaken, although such simulations are unfortunately computationally prohibitive at present. 

\subsection{Selecting individual HVCs from the simulation}\label{selsim}
We identified potential halo HVCs in three snapshots at 300, 400 and
600 Myr ($t$=10.3, 10.4 and 10.6 Gyr). First, all ``seed'' SPH particles
in the halo, satisfying $n_H>n_{H,trig}$ and $T$$<$3$\times$10$^4$ K,
were identified. Second, a
gas particle group finder was used to identify all SPH particles in
coherent regions in the halo, surrounding these ``seed'' particles, and
satisfying $n_H>n_{H,min}$. Third, only SPH particles in these regions
satisfying $T$$<$3$\times$10$^4$ K were retained. It is
found that with $n_{H,trig}$$\sim$10$^{-2}$cm$^{-3}$ and
$n_{H,min}$$\simeq$10$^{-3.5}$cm$^{-3}$
one identifies neutral or partly photo-ionized $T$$\sim$1-3$\times10^4$ K
gas in HVCs and satellite galaxies (SPH particles in the coherent regions of
$T$$\ge$3$\times$10$^4$ K typically have $T$$\sim$10$^6$ K and are
almost fully collisionally ionized), hence these density thresholds
were adopted. Subsequently,
the 7 satellite galaxies identified around this galaxy are removed on
the basis of these systems containing (1) gas of high central density
($n_H$$\ga$1-10 cm$^{-3}$), (2) stars and (3) dark matter. We note that none of these 7 satellite galaxies are accompanied by a stripped tail of gas analogous to the Magellanic Stream in any of the 3 snapshots. Indeed, there is no correlation between the location in position-velocity space of the satellites and the selected condensed clouds. We conclude from this that though it is possible that some of the condensed clouds we see were at one point associated with a small satellites, there is no evidence for halo clouds having been stripped from such satellites recently. We now describe the characteristics of these simulated halo clouds.

\subsection{Physical characteristics of simulated halo clouds}\label{physchar}

\begin{figure*}
\begin{center}
\includegraphics[scale=.66, angle=90]{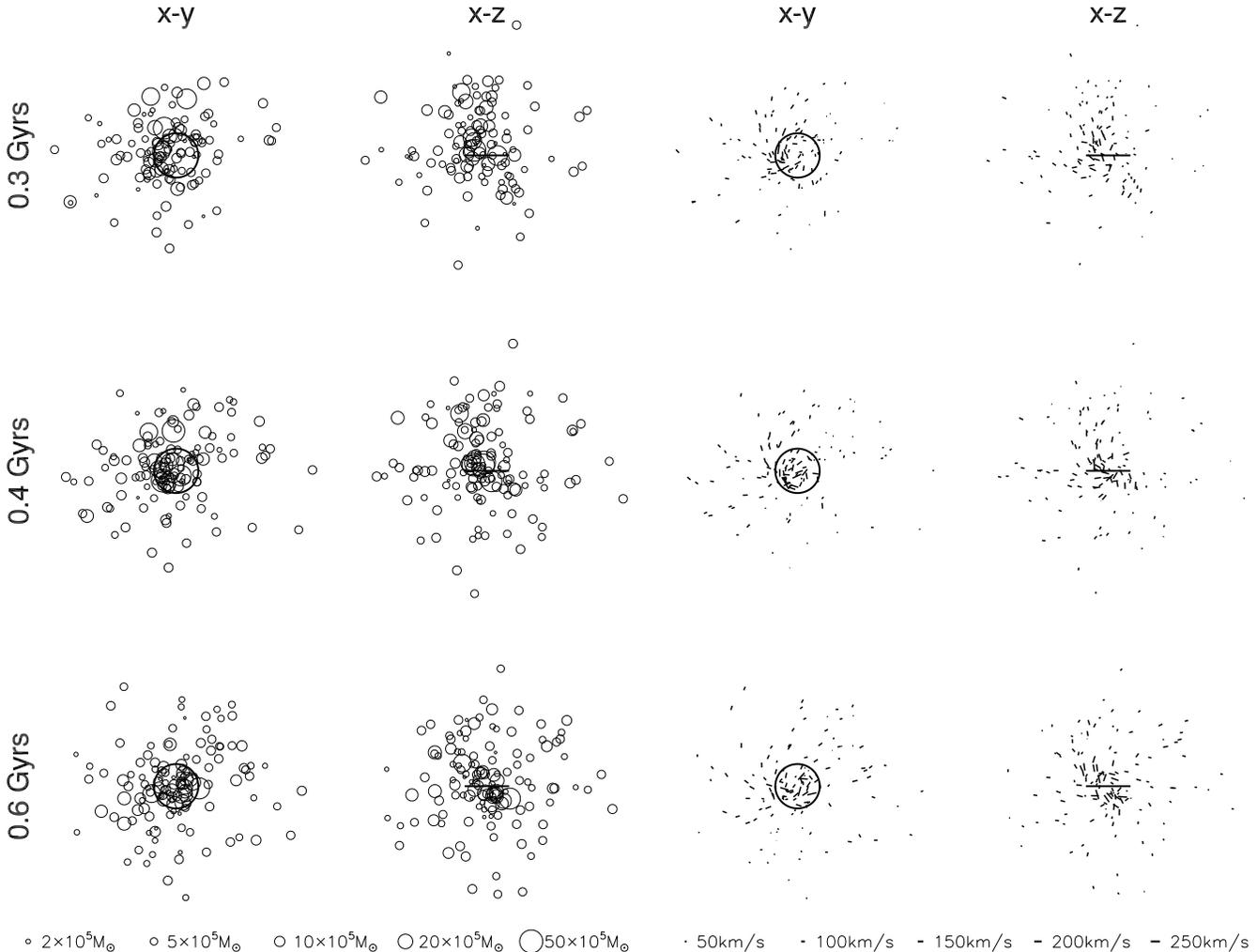}
\caption{The distribution of simulated HVCs in the three snapshots, showing HVC mass (left) and speed (right). Each snapshot is projected onto the x-y and y-z planes, and shows a ring of radius 15 kpc centered in the galactic plane for scale and orientation. Symbol area scales with the mass of the clouds at left, and symbol length scales with the speed at right.}
\label{posvel}
\end{center}
\end{figure*}

\begin{figure}
\begin{center}
\includegraphics[scale=.42, angle=0]{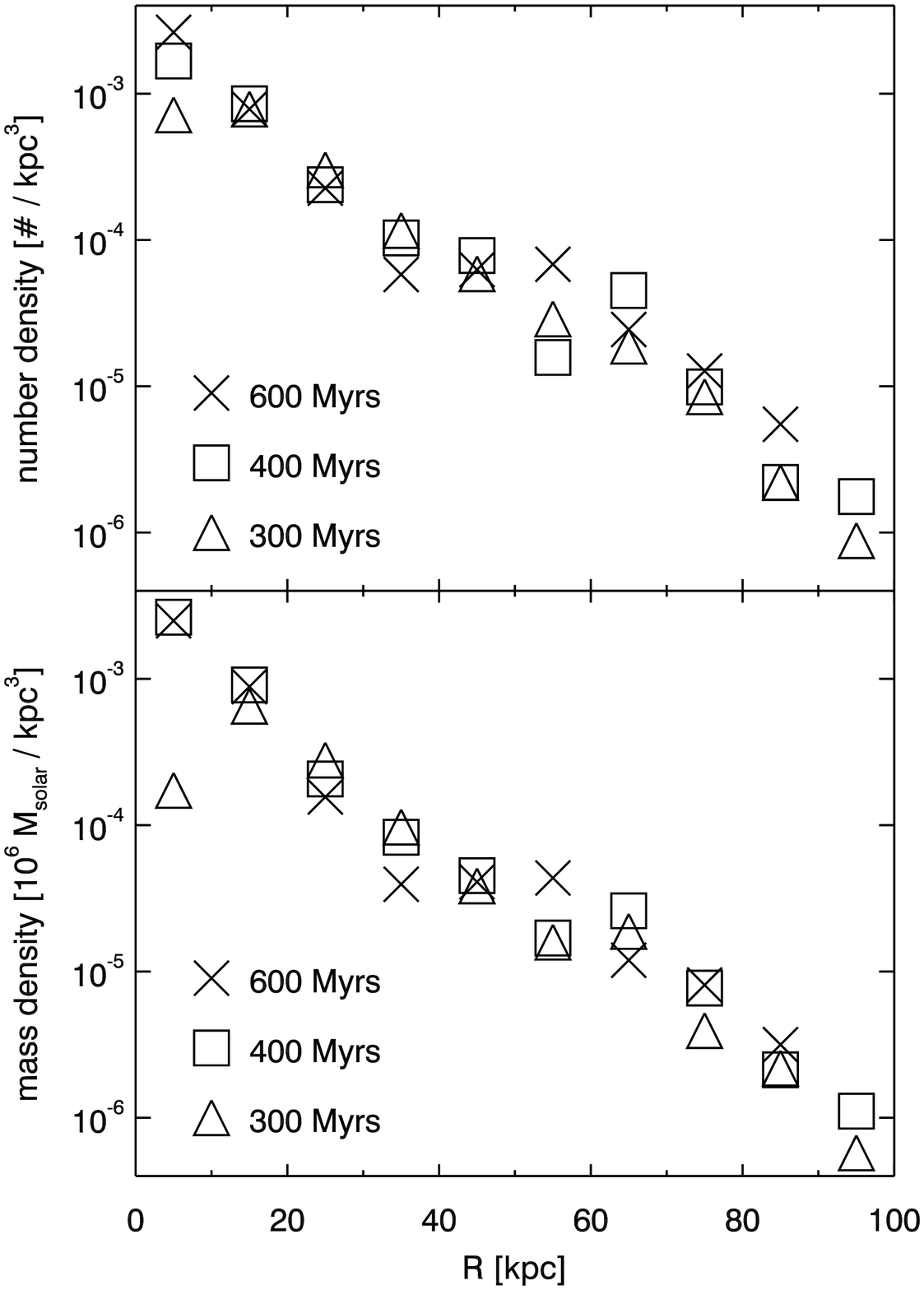}
\caption{The number density (top) and mass density (bottom) profiles for the simulated HVCs with galactocentric radius, in 10 kpc bins. Each of the snapshots is plotted. There are no clouds between 90 and 100 kpc in the 600 Myr snapshot, as compared to 1 and 3 clouds in the 300 and 400 Myr  snapshots respectively.}
\label{densprof}
\end{center}
\end{figure}

The three snapshots in the simulation (300, 400 and 600 Myrs) have 113, 128 and 130 identified halo clouds, respectively (hearafter, when describing a characteristic of the simulation, three quantities in a succession will refer to these three snapshots respectively). \citet{SL06} describes the mass distribution of the clouds in the simulation, which is consistent from snapshot to snapshot. Simulated halo cloud masses range from $10^5 M_\odot$ to $5 \times 10^6 M_\odot$, with total masses ranging from $8.8 \times 10^7 M_\odot$ to $10.7 \times 10^7 M_\odot$. Note that in the simulation of SL06, the highest resolution was achieved only in the region out to 50 kpc galactocentric radius, as compared to 100 kpc for the present simulation. Compared to the former simulation the number of warm clouds increased by about 50\%, but the total mass in warm clouds by less than 25\%. Also note that this total mass is much lower than the $\sim 2 \times 10^{10} M_\odot$ proposed in MB04. Figure \ref{posvel} shows the distribution of halo clouds in the three snapshots, indicating both masses and velocity vectors.  Note that the halo clouds in each snapshot are not independent; a typical cloud moving at 100 \kms will only traverse 30 kpc from the first to last snapshot, so large structures that are consistent from snapshot to snapshot may indeed be related. The cloud distributions are roughly spherically symmetric and have an overall sense of infall. Clouds tend to have velocity vectors similar to their neighbors and show large-scale (up to 30 kpc) inflow structures. This is worthy of note as large, coherent structures in HVCs are sometimes cited as qualitative evidence of a satellite-accretion origin. Figure \ref{densprof} shows the number density and mass density profiles of each of the snapshots. We note that the 300 Myr snapshot has a noticeably lower number and mass density of clouds near the disk. This effect is consistent with not enough time having elapsed since the beginning of the highest resolution simulation for the most massive clouds to have formed and fallen into the center of the potential. The 400 Myr and 600 Myr snapshots have very similar density structures, indicating a converged and continuous halo cloud lifecycle. 

\begin{figure}
\includegraphics[scale=.42, angle=0]{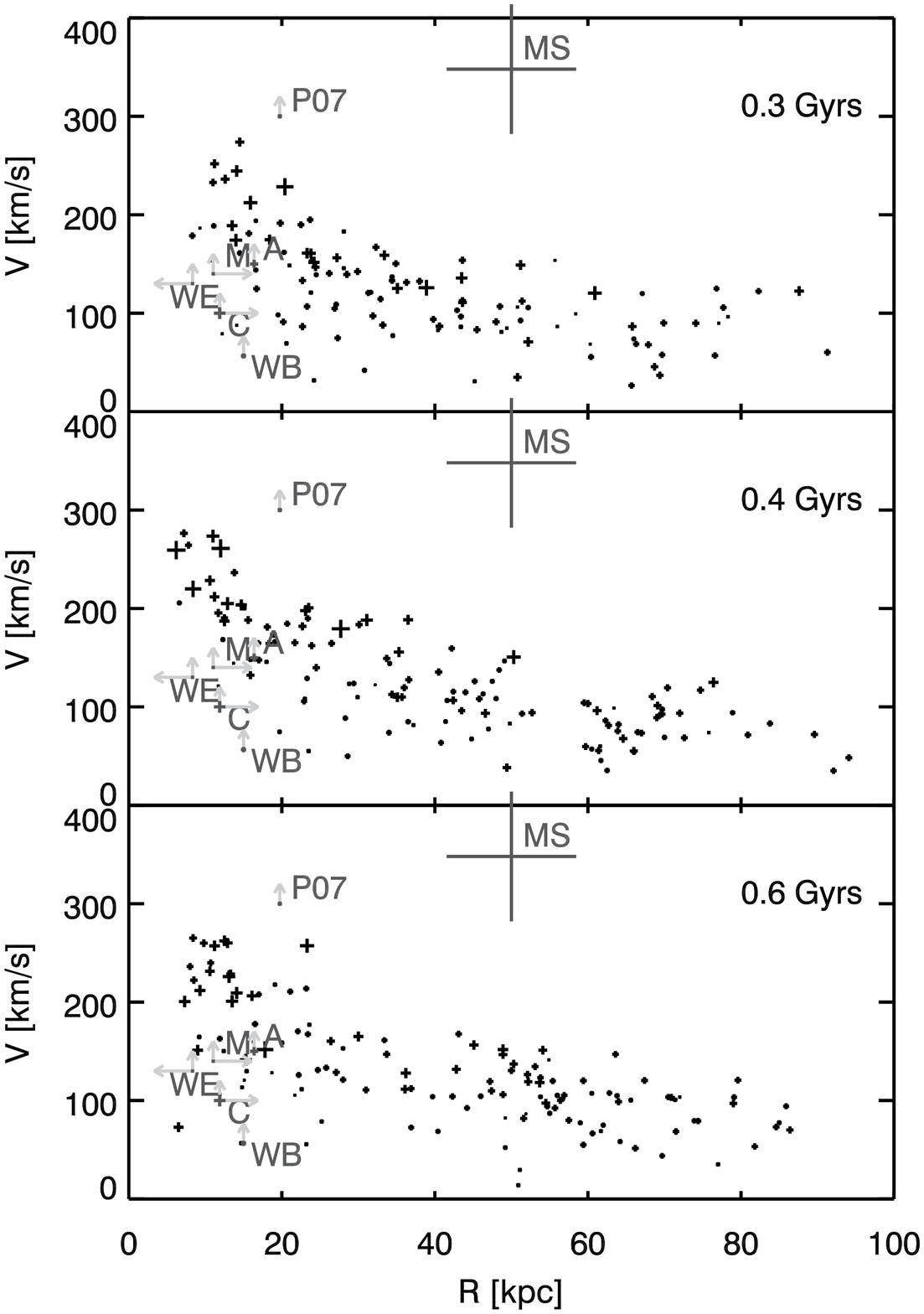}
\caption{Three plots of HVC galactocentric radius versus speed (with respect to the center of mass of the Galaxy) of the three snapshots in the S-L06 simulation. The size of the symbol scales as the square-root of the mass of the cloud. Clear trends of velocity with radius, mass with radius and velocity with mass at a fixed radius are evident. Six observed clouds and complexes that have distance constraints are over-plotted (data from \citet{vww04}, \citet{WVW2001} and \citet{Thom2006}), with arrows indicating where values are only upper or lower limits in distance or lower limits in velocity.}
\label{radvel}
\end{figure}

In Figure \ref {radvel} we plot the distribution of speed in the galaxy rest frame with respect to galactocentric radius of the halo clouds for each of the three snapshots. The most striking feature of this plot is the increasing velocity of clouds with decreasing radius, consistent over time, which is also visible in Figure \ref{posvel}. This should not be a terribly surprising result, as clouds deeper in a galactic potential will typically have more kinetic energy, but the simple idea that more distant HVCs may have velocities that are less extreme than other HVCs has not been much addressed in the study of HVCs to date. Also evident in Figure \ref {radvel} is that at a given distance, more massive clouds typically move faster. This effect consistent with the supposition that drag from the gaseous halo on halo clouds has a significant kinematic effect, and that more massive clouds typically have higher column densities.

\subsection{Observing simulated halo clouds}

To make comparisons between the simulated halo clouds to observed HVCs, it is necessary to ``observe'' the simulated clouds to determine the angular position, flux and radial velocity of each cloud. We do this by positioning a ``sun'' in the plane of the simulated galactic disk, 8.5 kpc from the galactic center and measuring each of those quantities for each cloud in each simulation snapshot. Fluxes are determined by assuming the gas in HVCs is 70\% hydrogen by mass (consistent with Big Bang nucleosynthesis), and that the clouds are optically transparent to 21-cm radiation. This transparency assumption is reasonable as HVCs have a peak brightness temperature of a few K and a spin temperature upwards of 1000 K (see \citet{KH88} for a discussion of the details of HI radiative transfer). There remains an unknown neutral fraction in the clouds, as the simulation does not carefully track the ionization of the halo clouds, which scales linearly with flux. Radial velocities of halo clouds can be measured including (LSR frame) or excluding (GSR frame) the effect of galactic rotation. There is an ambiguity in the angular position of the ``sun'' around the galactic center when taking these ``observations''. To overcome this, we measure each parameter at 8 solar positions:  R = 8.5 kpc,  $\phi = n \pi/4$, $ n= 1, 2, 3... 8$, in galactic cylindrical coordinates, applying the velocity cuts as described in section \ref{selhvc} for each point of observation. When reporting an ``observed'' value we report the mean of the 8 ``observations'', with an error that is the standard deviation of these 8 ``observations''. We report this measurement for the 300 Myr, 400 Myr and 600 Myr snapshots sequentially. As an example, the number of simulated halo clouds that would be considered HVCs, given our adopted radial velocity cuts, are $56 \pm 7$, $67 \pm 13$ and $74 \pm 9$.

\section{Creating HVC complexes}\label{cplxing}

To compare simulated halo clouds and observed HVCs it behooves us to group them into complexes in a consistent manner. The tendency for HVCs to be near one another in angle-velocity space (i.e. complexes) should be easily captured by the simulation, even if the total number of smaller clouds within a complex may not. One difficulty in generating a consistent method for determining membership of a cloud to a complex is the differing limitations in sensitivity of the simulations and the observations. The WvWcat is essentially a flux-limited catalog - very low-mass clouds that are particularly nearby are as easily observed as distant, massive clouds. The simulation, by contrast, is limited in mass resolution. The sensitivity of the observations to resolve nearby, very low mass clouds may be the reason for the very large number of low-flux clouds in the WvWcat that do not have analogs in the simulation data. To eliminate as many of these observed `chaff' clouds as possible, we wish to determine the minimum flux an HVC can have in the WvWcat to have a mass and distance that can be captured by the simulation. For clouds with fluxes of up to 300 Jy \kms to have masses that can be resolved by the simulation, they must be at distances of at least $\sim$ 50 kpc. If these clouds did largely exist at these types of distances and beyond, they would contribute $> 4 \times 10^7 M_{\odot}$ of neutral hydrogen to halo. The observations in T04 and \citet{Westmeier2005} show that most of the mass of extragalactic HVCs ($\sim\left(3-4\right) \times 10^7 M_{\odot}$) exists within 50 kpc. Under the assumption that the Milky Way and Andromeda do not have qualitatively different HVC halo populations, clouds with fluxes below 300 Jy \kms cannot consistently be at distances greater than 50 kpc and therefore cannot be reproducible by the simulation. This assertion is consistent with H $\alpha$ observations that show many small clouds to have ionization indicating a distance less than 40 kpc (\citei{Putman03}, \citei{Tufte2004}). We therefore make a flux cut at 300 Jy \kms, excluding all lower flux clouds in the WvWcat from our comparison. Though this cut removes 77 \% of the individual HVCs in the WvWcat of interest, it removes only 5\% of the flux.

Now that we have reasonably comparable data sets, to determine which clouds should be associated into which complexes we define a `distance measure' in position-velocity space between clouds:
\beq
D = \sqrt{\Theta^2 + f^2\left(\delta v\right)^2}.
\eeq
Here, $\Theta$ is the angular distance between two clouds, $\delta v$ is the difference in GSR velocity between two clouds and $f$ is a conversion factor; $f$ parameterizes the significance we ascribe to the angle subtended by two clouds versus their difference in velocity in determining whether they are members of the same complex. We choose $f = 0.5^{\circ} / $ \kms, broadly consistent with the clustering in historical HVC complexes. We place two clouds in the same complex if their D is less than some number, $D_0$. A cloud with no neighbors is considered a complex of its own. The overall complex position and velocity are determined by a flux-weighted average of the constituent clouds. 

This formulation leaves us with a free parameter, $D_0$. As $D_0$ is increased the number of complexes decreases roughly linearly for reasonable values of $D_0$ in the simulated and WvWcat data sets, thus the data do not offer a specific scale at which cloud clustering takes place. We wish to choose $D_0$ such that we maximize the identification of true complexes and minimize the identification of complexes that are the result of coincidental cloud superposition. To that end we scramble all of the positions (l and b) and velocities ($v_{\rm obs}$) of the simulated clouds within each snapshot, as well as in the WvWcat data, such that all of the coherent angular and velocity structures are lost, while maintaining the distribution of each of these parameters individually. We then run the clustering algorithm on the scrambled and unscrambled data sets for all values of $D_0$. We assume that any complexes generated in the scrambled data sets are spurious and we find the average $D_0$ at which the difference in clustering, as parameterized by the number of complexes, is greatest between the true and scrambled data sets. This maximum occurs at $D_0 \simeq 25^{\circ}$, for each of the snapshots as well as for the WvWcat data. We adopt this value for generating cloud complexes hereafter, and refer to complexes from the S-L06 halo clouds as `simulated complexes' and complexes from the WvWcat HVCs as `observed complexes'. 

\section{Comparison Between Simulated and Observed Clouds}\label{anal}

\subsection{Milky Way HVC Complexes Compared to Simulated Cloud Complexes}

\begin{figure}
\includegraphics[scale=.42, angle=0]{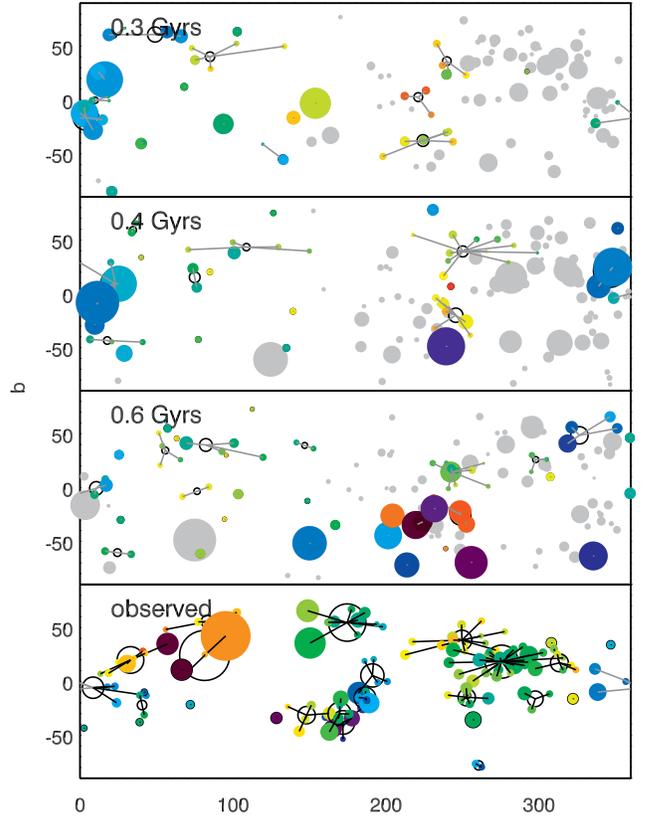}
\caption{Four plots of HVC complex distribution on the Galactic sky. The first three are from the S-L06 simulations and the final plot is of the observed HVC complexes. Filled circles represent HVCs with the linear size of the circle scaling with the cube-root of the HI flux, as HI flux covers more than 3 orders of magnitude. In the simulation plots, clouds are gray if they do not fulfill the velocity selection criteria: $|v_{\rm dev}| > 60$ \kms and $|v_{\rm LSR}| > 90$ \kms. Clouds are colored by their velocities, and we assume an neutral fraction of $70\%$ in the simulation plots. Lines connect clouds to the flux-weighted centers of the complexes to which they belong. Complexes are indicated by an empty circle, with fluxes scaling the same way with circle size.}
\label{cplxs}
\end{figure}

\begin{figure*}
\begin{center}
\includegraphics[scale=.60, angle=90]{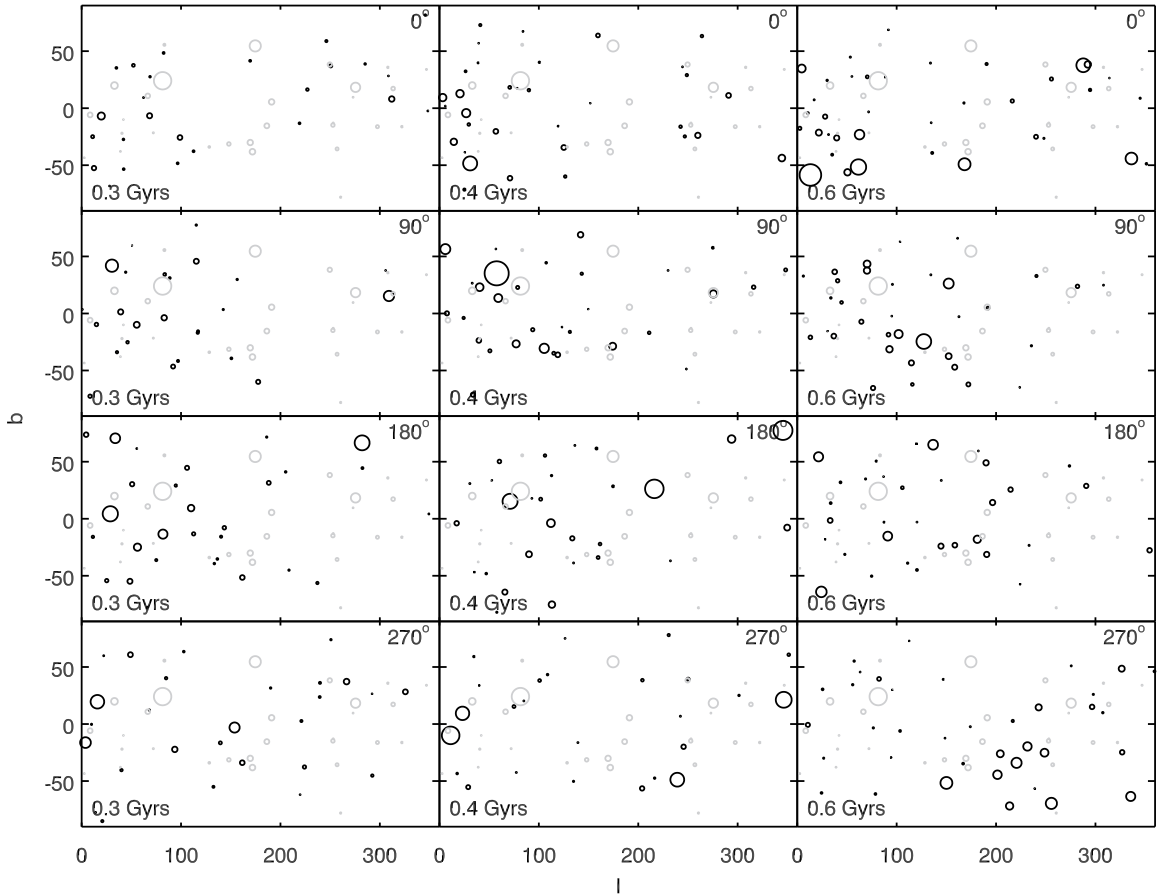}
\caption{Twelve plots of Galactic l versus Galactic b for simulated HVC complexes. Columns from left to right are from each sequential snapshot (0.3, 0.4 and 0.6 Gyrs into the high-resolution simulation), and rows are from positions rotated by $90^{\circ}$ around the solar circle. The symbol area scales with flux. Dark symbols represent simulated complexes and light symbols represent observed Galactic complexes, for comparison. Symbols are scaled using 70\% neutral fraction. Note that the distribution of simulated complexes varies significantly with position of the Sun within the solar circle, but that the simulated complex distributions are all qualitatively similar to the observed complex distribution.}
\label{pospos}
\end{center}
\end{figure*}

First we wish to compare the number and angular size of observed HVC complexes to the simulated complexes that fit our velocity selection criteria as outlined in \S \ref{obs}. Figure \ref{cplxs} shows each of the snapshots as observed from the Sun along with the observed HVC data. The number of complexes is similar, $20.7 \pm 1.5$, $26.1 \pm 3.3$ and $29.0 \pm 2.0$, as compared to 29 in the WvWcat. Some subtle evolution exists in the number of clouds and complexes in the simulation data, consistent with the halo requiring more than 300 Myrs to reach an equilibrium in the condensation of HVCs after the simulation has been run at high-resolution. Note that the simulated complexes have rather a random distribution on the sky, similar to observed complexes. Simulated complexes also range in angular size up to about $60^{\circ}$, consistent with range of angular sizes in the observed complexes.  In Figure \ref{pospos} we show simulated complexes from four vantage points at the solar circle for each snapshot, demonstrating the variation in the distribution of complexes as a function of solar position. This overall agreement between simulation and observation is consistent with earlier results in \citet{Connors2006}.

\begin{figure}
\includegraphics[scale=.43, angle=0]{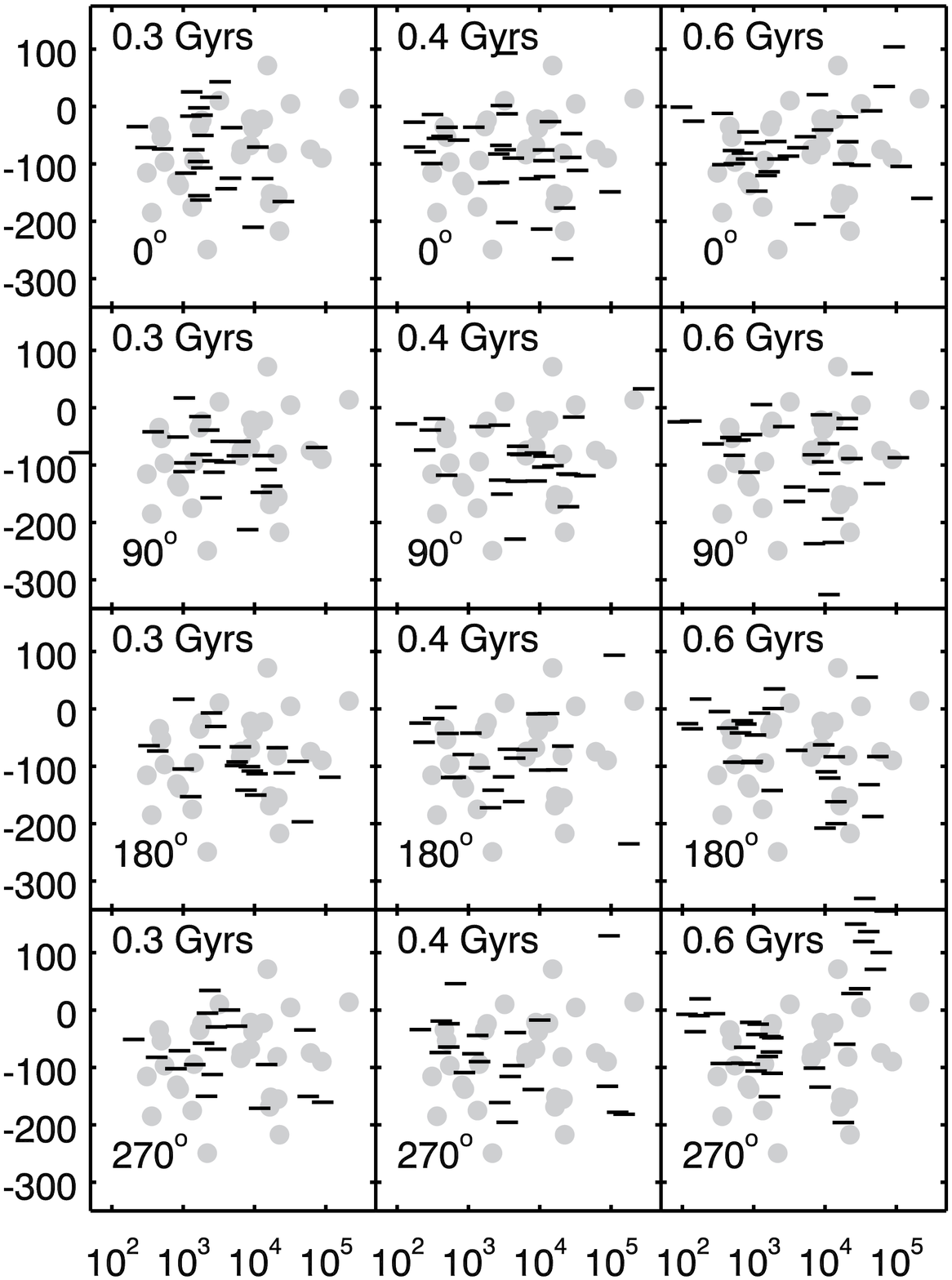}
\caption{Twelve plots of GSR velocity versus HI flux for simulated HVC complexes and observed HVC complexes. In each plot the observed HVC complexes are depicted by gray filled circles for comparison. The dashes represent simulated complexes, with the right side of the dash representing 100\% neutral and the left side of the dash representing 50\% neutral. From left to right the plots are of 300, 400 and 600 Myrs into the simulation. From top to bottom the plots represent viewpoints of $0^{\circ}$, $90^{\circ}$, $180^{\circ}$ and $270^{\circ}$ around the solar circle.}
\label{velflux}
\end{figure}

We also wish to compare the velocities and fluxes of the simulated complexes and observed HVC complexes. Figure \ref{velflux} shows the fluxes and velocities for the simulated complexes in each of the snapshots from four vantage points around the solar circle, along with the observed HVC complex data. The total maximum HI flux in the simulated complexes are $4.5 \pm 1.9$, $10.0 \pm 4.6$ and $8.3 \pm 2.9$ $\times 10^5$ Jy km/s, as compared to 5.7 $\times 10^5$ Jy km/s in the compared sample of observed HVC complexes. Note that the maximum flux comes from the assumption that all of the simulated cloud mass is neutral, therefore if only a small fraction ($\sim 30\%$) of the simulated halo clouds were neutral, simulation and observation would have significant discrepancy in HI flux. The range in total flux within a single snapshot is due to the fact that $\sim 50\%$ of the overall flux comes from just a few simulated complexes closer than 10 kpc; as the observation point is rotated around the Galaxy, the distance to these complexes changes, changing the overall flux. This dependence upon only a few large, local clouds, though consistent with observations, diminishes the usefulness of flux as a measure of the accuracy of the simulations. The average velocities (GSR) are also similar: $-81 \pm 7$, $-82 \pm 6$ and $-65 \pm 24$ \kms, as compared to $-85$ \kms in the observed HVC complex dataset. The standard deviation of the distribution of simulated cloud complex velocities is $56 \pm 4$, $68 \pm 10$ and $74 \pm 25$ \kms, as compared to 75 km/s in the observed complex data set. These distributions of observables are consistent with the hypothesis that the condensed halo clouds in the S-L06 simulations are analogous to the Milky Way's population of HVCs.

\subsection{Andromeda's HVCs Compared with Simulated Cloud Complexes}

Here we compare observed quantities in the T04 Andromeda dataset to our simulated data set. We make an analogy here between the observed extragalactic HVCs around Andromeda and simulated complexes, as the masses are similar. T04 find that there is $\sim\left(3-4\right) \times 10^7 M_{\odot}$ in HI mass around Andromeda; we find 6.1, 7.5 and 6.8 $\times 10^7 M_{\odot}$ in our 3 simulation snapshots of 300, 400 and 600 Myrs. This implies an neutral fraction ranging from 40\% to 65\% for the simulations to be consistent with the observations. The standard deviation of the projected velocities in the T04 Andromeda dataset is 126 km/s, although limiting the sample to objects with masses greater than $5 \times 10^5 M_{\odot}$, where we expect the sample to be complete, reduces this to 96 km/s. Also note that this velocity dispersion only pertains to the qualitatively-defined ``objects'' in the T04 dataset, which excludes diffuse gas closer to the Andromeda disk. The simulation snapshots, once projected to mimic the Andromeda viewing angle, have velocity standard deviations of 63, 71 and 89 \kms. 

The spatial distribution of the halo clouds is one of the most interesting quantities to pursue in this comparison, as the Andromeda data set yields a minimum galactocentric distance for all observed extragalactic HVCs, a quantity not afforded by the Galactic data set. It is, unfortunately, a rather difficult quantity to extract, as the scale of distribution of the halo clouds is very sensitive to the sensitivity of the observations to small clouds, which is difficult to accurately characterize in these observations. We can conclude that both the Andromeda data set and the simulation snapshots have detectable HVCs out to projected galactocentric radii of 50 kpc, and lack clouds with $M > 10^6 M_\odot$ outside 50 kpc \citep{Westmeier2005}. A more detailed analysis of the scale-length and HVC density profile, including lower mass clouds, will have to wait for deeper, higher resolution observations.

\section{Analysis \& Discussion}\label{disc}

In this section we discuss the significance of ascribing a cooling-formation origin consistent with the S-L06 simulations to the observed HVC population for a few different HVC observables. We also discuss a few ways to observationally disprove the hypothesis that the bulk of observed HVCs have this origin. 

\subsection{The Speed-Radius Domain}

Consistency between simulations and observations can be further examined through investigation of the speed-radius relationship for both populations. The distribution that the simulated clouds present (see Figure \ref{radvel}) has a very clear variation of Galactic rest-frame speed with galactocentric radius. The vast majority of the observed HVCs with distance constraints are consistent with the simulated HVCs. Only one observed complex in the Galactic sky with a known distance significantly strays from this region: the Magellanic Stream. This discrepancy, of course, does not contradict the hypothesis that the rest of the HVCs are generated by a mechanism similar to the one that generated the S-L06 halo clouds; the Magellanic Stream was excluded from the analysis for the specific reason that is has a well-known, non-cooling origin. The deviation of the Magellanic Stream from this area does indicate, however, that HVCs that are generated from satellite accretion may occupy a different part of the radius-velocity parameter space and that it therefore may be possible to distinguish the origin of an individual cloud if these values can be measured.

It is worth noting that the very high-velocity cloud (VHVC) 160.7-44.8-333 (annotated P07 in Figure \ref{radvel}) has been suggested to have a galactocentric distance of $\sim 20$ kpc (\citei{WVW2001}; \citei{Peek07}). At a minimum velocity of $300$ \kms GSR, it is also marginally outside of the domain of clouds that are generated in the S-L06 simulations. Further distance limits on VHVCs may help to show whether these clouds are broadly inconsistent with cooling formation scenarios. Observations of VHVCs are particularly important as they have very large minimum Galactic rest frame speeds, and are therefore plausibly excludable from the cooling-formation-origin domain. 

\subsection{HVCs at Low Velocity}

A crucial byproduct of the analysis of simulated condensed clouds as observable HVCs is that \emph{nearly half} of all simulated clouds would not satisfy the observational velocity criteria for HVCs. Instead, these clouds would be construed as lower-velocity (and therefore nearby) gas. This is to say that if the bulk of HVCs are generated by processes similar to those in the S-L06 simulations we are `missing' as many HVCs as we observe and thus underestimate the total scale of HVCs by a factor of two. This effect increases with greater radius from the galaxy - the average velocity of clouds decreases with radius (see Figure \ref{radvel}), such that more distant clouds will be more often confused with local gas. The clouds with low radial velocity are startlingly asymmetrical. The ratio of the number of condensed clouds above $|b|$ of 50$^\circ$ with $-50$ \kms $< V_{GSR} < 0$ \kms to clouds with $0$ \kms $< V_{GSR} < 50$ \kms is $7 \pm 4$, $9 \pm 3$ and $11 \pm 3$ for each of the snapshots. If these clouds could be observationally deciphered from nearby LVCs and IVCs, this asymmetry could be a powerful test of this model for HVC formation. It may be possible in the future to disentangle such clouds from local gas via their morphological characteristics, absorption to stars, low metallicity and very low dust-to-gas ratio. 

\subsection{CHVC hypotheses}

Compact high-velocity clouds (CHVCs), are HVCs that are smaller than about 2$^\circ$ in projected size and are not directly linked with other HVCs in the sky. CHVCs have received significant attention in recent years (e.g. \citei{BB1999}) under the assumption that some or all of them are analogous to the observed large complexes of HVCs but at much greater distances. Were this true, CHVCs could dominate the mass of Galactic HVCs, and would be very important to understanding the structure of the HVCs and the Galactic halo. This assumption hinges on the assumption that the mass of an HVC complex is relatively independent of distance from the Galaxy. If HVCs have a mass-distance distribution similar to the simulated distribution, this assumption would prove false. The most massive halo clouds in the S-L06 simulations are typically closest to the disk, and they do not have analogs 50 to 100 kpc from the galaxy. In addition to this, the simulated velocity distribution of local halo clouds ($< 30$ kpc) is not analogous to that of more distant halo clouds; distant halo clouds move slower in the S-L06 simulations. Thus if some CHVCs are indeed distant objects (and the cooling-formation predictions are correct) there will exist a trend toward lower velocities in CHVCs. CHVC catalogs do not show a lower velocity dispersion than catalogs of HVCs \citep{putman2002}, but such a discrepancy may be hard to detect as small, local clouds could easily masquerade as distant CHVCs (thus contaminating the sample) and disk gas may obscure a large fraction of the population of slow-moving, distant CHVCs. Conversely, it is possible to limit the contribution to the HVC population from cooling-formation HVCs if CHVCs with $V_{GSR} > 200$ \kms can be shown to be distant ($R > 50$ kpc). This is to say the velocity distribution of distant clouds may be a strong test of this cooling-formation scenario and the SL06 simulations.

\subsection{The HVC Accretion Rate}\label{accretion}

\begin{figure}
\includegraphics[scale=.32, angle=90]{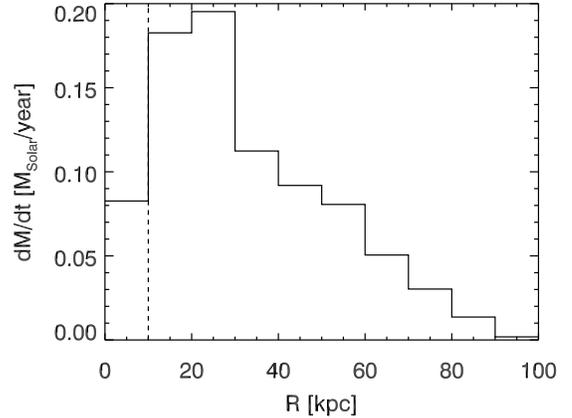}
\caption{The accretion rate onto the simulated galaxy as a function of radius, as defined by the equation $\dot{M}\left( R\right) = \sum_{i=1}^{i=n\left(R\right)}\frac{M_{i} \vec{V_{i}} \cdot \left(-\hat{r_i}\right)}{dR}$. The plot shows the average across all three snapshots. Inside the dashed line the parameterization of the accretion rate with radius is no longer meaningful as clouds are merging with the disk.}
\label{accr}
\end{figure}

In this section we compare the accretion rate in the S-L06 simulations, and the source of that accretion, to the accretion necessary to fuel the ongoing star formation observed in our Galaxy. In the S-L06 simulations, material cools out of the halo at different radii, allowing us to parameterize the halo clouds accretion rate onto the galaxy as a function of radius in the simulations. At R=100 kpc the accretion rate is zero and the rate increases with decreasing R. We define the halo clouds accretion rate at a given galacto-centric radius by
\beq
\dot{M}\left( R\right) = \sum_{i=1}^{i=n\left(R\right)}\frac{M_{i} \vec{V_{i}} \cdot \left(-\hat{r_i}\right)}{dR},
\eeq
where $M_i$ is the mass of a cloud, $\vec{V_{i}}$ is a cloud velocity and $\hat{r_i}$ is the radial unit vector at the cloud. $n\left(R\right)$ is the number of clouds that exist in a spherical shell from $R-dR/2$ to $R+dR/2$. $\dot{M}\left( R\right)$ is not a well-defined metric within 10 kpc of the galactic center, as some of the halo clouds will be removed from the system when they collide with the disk, so we exclude HVCs within 10 kpc from this analysis. We find that this halo clouds accretion rate, averaged over all three snapshots, can be easily fit by a line with $\dot{M}\left( 0\right) = 0.22 \pm 0.014 M_{\rm{\odot}}/\rm{year}$ and $d\dot{M}\left( R\right) /dR = -2.5 \times 10^{-3}  \pm 2 \times 10^{-4} M_{\rm{\odot}}/\rm{year} / \rm{kpc}$ (see Figure \ref{accr}). This disk accretion rate of 0.22 $M_{\rm{\odot}}/\rm{year}$ is consistent with the assertion that HVC accretion is not the sole source of fuel for Galactic star formation unless the accretion rate was higher at earlier in the lifetime of the Galaxy (see \citei{Putman06} and \citei{SL06}). The monotonic and linear increase in accretion rate with decreasing radius demonstrates that the cooling process that drives halo cloud formation and growth operates at all radii within the halo in the S-L06 simulations. This result may indicate that HVCs form and grow throughout our own Galactic halo as well. This is contrary to a simpler picture wherein all HVCs form at some typical distance in the halo and fall toward the Galactic disk.

\section{Conclusion}\label{conc}

The number of complexes as identified in our TreeSPH simulation and their angular distribution on the sky is consistent with the Milky Way HVCs, when the velocity selection effects and clustering effects are taken into account. The radial distribution of simulated condensed halo clouds is consistent with the T04 sample of extragalactic HVCs around Andromeda and, in particular, the lack of massive ($\sim 10^6 M_{\odot}$) clouds at large radii ($> 40$ kpc) in the simulation is consistent with observations of Andromeda which show a dearth of such clouds far from the disk. The flux and velocity distributions of simulated complexes are also consistent with observations of HVC complexes in the Milky Way and of extragalactic HVCs in Andromeda, and point to neutral fractions above $\sim30\%$. This lower limit is inconsistent with a ``tip-of-the-iceberg'' picture of HVCs, wherein HVCs are dominantly an ionized phenomenon and the observed HI is just a small fraction of the baryonic mass of the clouds. ÊThe physical and observational population characteristics of the simulated halo clouds are consistent over 300 Myrs (excepting the nearest, most massive HVCs in the first snapshot) and are broadly independent of the point of observation chosen on the solar circle. We have found that simulated halo material condenses into clouds from R = 10 kpc to R =100 kpc in the halo, and that these clouds have an overall accretion rate of $\sim 0.2 M_\odot /$ year. 

We have also shown that the speed-distance domain is populated by simulated halo clouds only in a specific region, and that all HVCs with known distances and unknown origins (i.e. non- Magellanic) reside in this region. This points to a possible method for discriminating HVC origin given HVC speed. The simulated clouds are not always identifiable as HVCs, as they may have low projected velocity. Indeed, we find that half of the simulated halo clouds do not show high enough projected velocities to be considered HVCs at all. Specifically, distant simulated halo clouds are typically less massive and slower moving than nearby simulated halo clouds which makes them harder to distinguish from local gas. This implies that the populations of distant CHVCs and nearby HVCs are physically different, if we assume that CHVCs and HVCs are formed through a similar cooling mechanism.

Observations using the Arecibo L-Band Feed Array will refine our understanding of both the Galactic HVC population with the GALFA-HI observing program (e.g. \citei{Stanimirovic2006} and \citei{Peek07}) and the distribution of HVCs around other galaxies with the AGES observing program (e.g. \citei{Auld2006}). AGES will generate maps with similar fidelity to those in T04, but for more galaxies out to $\sim 5$ Mpc. Instruments coming online now, such as the Allen Telescope Array, will allow us to map the HI HVCs in the vicinity of other nearby galaxies with unprecedented efficiency and further compare these cooling-formation simulations to observed systems. Indeed, the full ATA-350 will be able to make maps comparable to those in T04 out to $\sim 10$ Mpc in a single day's observation. In this way we may be able to determine whether there are characteristics of extragalactic HVC systems that can inform our understanding of the formation of these galaxies and the variation in the character of their baryonic halos. In the more distant future, the Square Kilometer Array will allow us to extend this analysis to non-zero redshift, probing the cooling history of galaxies toward the age of mergers. To match these observational efforts, more detailed and comprehensive simulations will be required. To this end, cosmological simulations of even higher resolution than the ones described in the paper, and with a more sophisticated treatment of UVB and stellar ionizing photon radiative transfer (e.g. \citei{RS-L07}) are underway.

The authors thank David Thilker and Bart Wakker for generously providing access to their HVC data sets. The authors would also like to thank Carl Heiles, Kathryn Peek, Evan Levine and Andrey Kravtsov for many helpful conversations and our referee, Ari Maller, for many useful comments. The TreeSPH simulations were performed on the SGI Itanium II facility provided by DCSC. The Dark Cosmology Centre is funded by the DNRF. The research of JEGP was is supported in part by NSF grant AST04-06987. 

\bibliographystyle{apj}

\end{document}